\documentclass[aps,twocolumn,superscriptaddress,longbibliography]{revtex4-1}

\usepackage[utf8]{inputenc}

\usepackage{graphicx,epsfig,amsfonts}
\usepackage{bm}
\usepackage{amsmath}
\usepackage{amssymb}
\usepackage{float}
\usepackage{xcolor}
\usepackage[colorlinks=true,citecolor=blue]{hyperref} % hyperreferences

\DeclareGraphicsExtensions{.png,.jpg,.eps}
\usepackage{xcolor}

\newcommand{\beq}{\begin{eqnarray}}
\newcommand{\eeq}{\end{eqnarray}}
\renewcommand{\H} {\mathcal{H}}

\begin{document}

\author{Maryam Khosravian}
\affiliation{Department of Applied Physics, Aalto University, 02150 Espoo, Finland}

\author{Elena Bascones}
\affiliation{Instituto de Ciencia de Materiales de Madrid (ICMM), Consejo Superior de Investigaciones Científicas (CSIC), Sor Juana In\'es de la Cruz 3, 28049 Madrid, Spain}

\author{Jose L. Lado}
\affiliation{Department of Applied Physics, Aalto University, 02150 Espoo, Finland}

\title{{Moir\'e-enabled topological superconductivity} in twisted bilayer graphene}

\begin{abstract}
Twisted van der Waals materials have risen as highly tunable platforms for
	realizing unconventional superconductivity. Here we demonstrate how a
	topological superconducting state can be driven in a twisted graphene
	multilayer at a twist angle of approximately 1.6 degrees proximitized
	to other 2D materials. We show that an encapsulated twisted bilayer
	subject to induced Rashba spin–orbit coupling, s-wave
	superconductivity, and exchange field generates a topological
	superconducting state enabled by the moiré pattern. We demonstrate the
	emergence of a variety of topological states with different Chern
	numbers, that are highly tunable through doping, strain, and bias
	voltage. Our proposal does not depend on fine-tuning the twist angle,
	but solely on the emergence of moir\'e minibands and is applicable for
	twist angles between 1.3 and 3 degrees. Our results establish the
	potential of twisted graphene bilayers to create topological
	superconductivity without requiring ultraflat dispersions.
\end{abstract}

\date{\today}

\maketitle

\section{Introduction}

The two-dimensional nature of van der Waals materials provides a unique playground
to engineer emergent states\cite{Liu2016,Novoselov2016,Andrei2021}. Among others, their 2D nature 
allows to induce ferromagnetic exchange \cite{PhysRevB.82.161414,PhysRevB.97.085401,Zollner2019,Wolf2021,Chen2022}, superconductivity and spin-orbit coupling (SOC)\cite{Wang2020,Wang2019, Island2019, Naimer2023, Yang2017}
through proximity to other 2D materials. This versatility 
increases considerably the range of phenomena that can be explored in
van der Waals materials. In particular, the combination of Rashba SOC, Zeeman splitting and superconductivity in van der Waals heterostructures can give rise to topological superconductivity\cite{Beenakker2013,Alicea2012,10.21468/SciPostPhys.14.4.075,Mnard2017,Kezilebieke2020}. These ingredients, difficult to find in a single material, can be engineered in a van der Waals heterostructure through the appropriate combination of 2D materials. 
The demonstration of topological superconductivity in CrBr$_3$/NbSe$_2$ heterostructures\cite{Kezilebieke2020}
featuring moir\'e effects\cite{kezilebieke2022moire} opens the possibility to use the interplay of moir\'e
and impurities to probe unconventional quantum states\cite{khosravian2022impurity,2022arXiv221101038S,PhysRevMaterials.3.084003,PhysRevB.99.245118,2023arXiv230403018B}. 
Bulk and edge modes can be observed by tunneling through the insulator\cite{Kezilebieke2020, kezilebieke2022moire}, demonstrating that indeed edge modes can be observed even in the presence of a monolayer insulator between the tip and sample.

\begin{figure}[t!]
    \centering
   \includegraphics[width=\columnwidth]{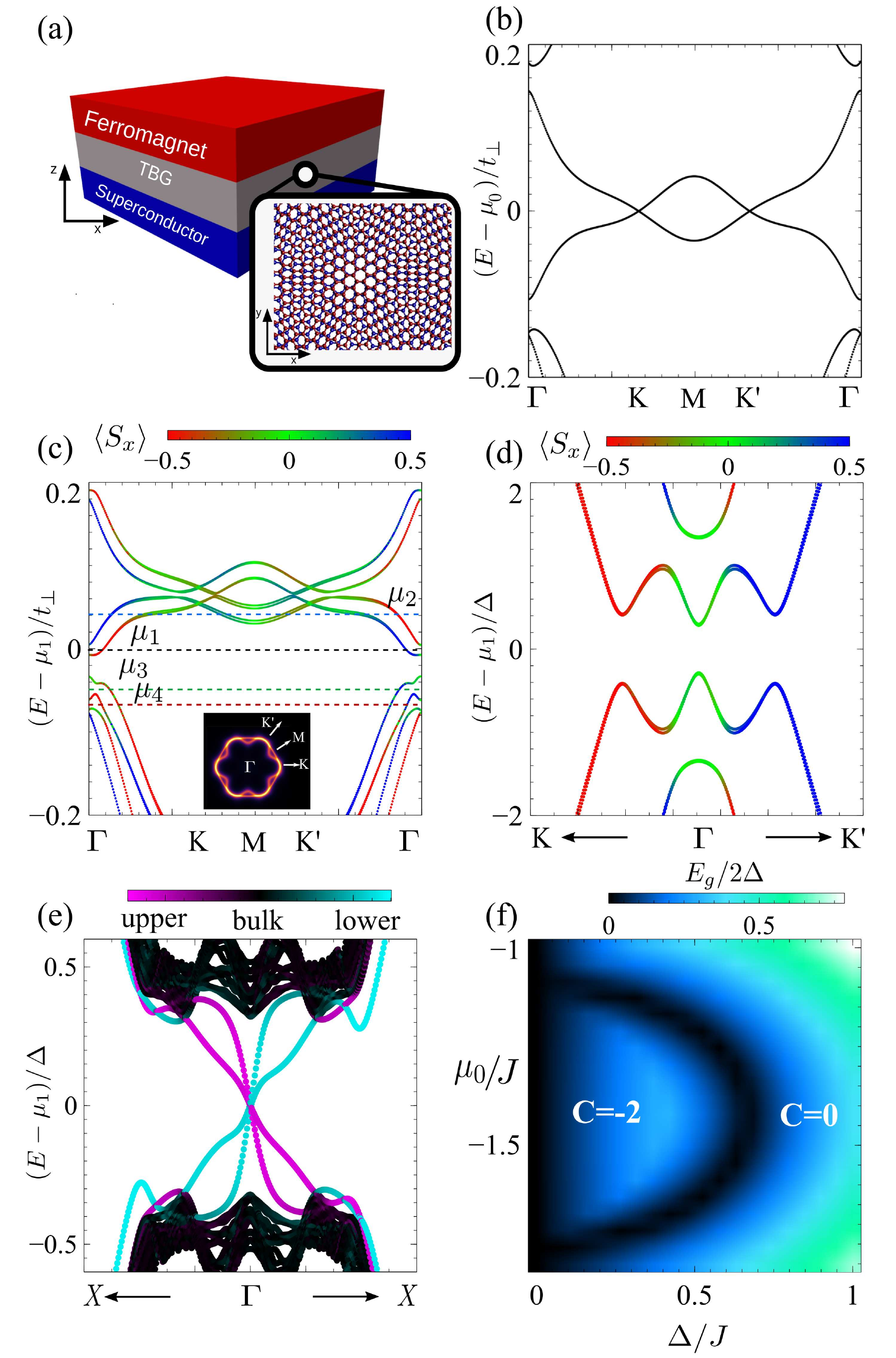}
    \caption{(a) Schematic of TBG  within ferromagnetic and superconductor layers. (b) Flat bands in TBG with a twist angle of $\alpha = 1.6 ^\circ$, without exchange and Rashba SOC. (c) Same as (b)  but with the induced exchange field and Rashba SOC used in the calculation and center at chemical potential $\mu_{1}$, corresponding to $\mu_0=-1.5$J. The color in (c) represents the expectation value of the in-plane spin operator $S_x$ in a given valley. Dashed lines mark the chemical potentials at which the Chern number is calculated. Inset: Fermi surfaces at $\mu_{1}$. (d) Zoom of the bands in (c) at $\mu_1$  when a superconducting order parameter $\Delta=0.4 J$ is induced in the lower layer of the TBG. (e) Energy states of a ribbon of width $N_r=100$ moir\'e unit cells, using the parameters in (d) showing evidence of $C=-2$ topological superconductivity. The color scale gives the projection of the states along $y$ direction. Cyan and magenta colors indicate that the state is localized at each of the edges of the ribbon. (e) Energy gap as a function of $\mu_0$ and $\Delta$. The gap vanishes at the transition between $C=-2$ and $C=0$ states. 
   } 
    \label{fig:fig1}
\end{figure}

Moir\'e patterns in twisted graphene multilayers
allow engineering 
quantum states including
superconductivity and correlated insulators\cite{Cao20181,Cao2018, Yankowitz2019, Lu2019, Chen2019,Arora2020,tripletCao}, orbital ferromagnetism\cite{Sharpe2019,Lu2019, Xie2021} and electronic cascades \cite{Wong2020,Zondiner2020, Saito2021, Datta2023}. 
 In twisted bilayer graphene (TBG) the moir\'e pattern opens gaps between low-energy narrow bands and high-energy wide bands, giving rise to a parabolic dispersion close to the $\Gamma$ and the M points\cite{dosSantos2007,SuarezMorell2010,Bistritzer2011,Fang2016}. The large unit cell in small twist angle TBG allows gate-doping up to several electron/holes per moir\'e unit cell, making it possible to probe a doping range spanning the full flat bands and partially the higher energy bands. The TBG bandstructure is highly sensitive to external parameters such as pressure\cite{Yankowitz2018,Carr2018},  strain\cite{Kerelsky2019,Bi2019,Sinner2022}, induced spin-orbit coupling (SOC)\cite{David2019} and electric field\cite{GonzalezArraga2017,Po2018,Carr2019}. The intrinsic correlation effects and the superconductivity, which originate in the flatness of the narrow bands, are restricted to angles very close to the magic value $\theta\sim 1.1^\circ$.
 However, the potential to engineer tunable states from parabolic dispersion at $\Gamma$ both in low and high energy bands has remained relatively unexplored.

Here we demonstrate that a highly tunable topological superconducting state can be driven in a twisted graphene multilayer with TBG stacked between ferromagnetic and superconducting layers. 
The mechanism proposed here does not require tuning the twist angle to the flat band value, being generically applicable to twisted bilayers between 1.3 and 3 degrees. 
{The upper limit for twist angle ($<3^\circ$) ensures a large unit cell and the possibility of achieving large doping with a gate electrode. The lower limit ($>1.3^\circ$) ensures that the intrinsic correlated and superconducting states of magic-angle TBG  are not present and therefore do not interfere with the mechanism for topological superconductivity proposed here. In this range of twist angles, the band structure used in our work is a good starting point.}
Rashba SOC, ferromagnetic exchange and an s-wave superconductivity are induced in TBG through the proximity to the other 2D materials. 
The mini-gaps driven by the moir\'e pattern enable the appearance of topological superconductivity.
Specifically, twisted bilayer graphene allows generating topological superconducting states with 
Chern number up to $|C|=6$, tunable with doping, strain, and electric field.

\section{Model}
We consider a heterostructure with twisted bilayer graphene encapsulated between a 
ferromagnetic insulator and an s-wave superconducting layer, 
Fig.~\ref{fig:fig1}(a).
Breaking of the mirror symmetry at the interfaces 
between the different materials generates Rashba SOC in TBG. 
{Although graphene has relatively weak SOC \cite{Yao2007}, 
spin-orbit coupling can get greatly enhanced via proximity effect to a substrate\cite{Avsar2020,Guimares2014, Wang2016, Island2019, Wang2019-hall, Wang2020}, which is the strategy we use here. For instance, transition metal dichalcogenides (TMDs) have been predicted to induce substantial Rashba SOC up to 15 meV 
%and spin–valley SOC around 1.5 meV 
in an adjacent graphene layer \cite{Wang2019}.} Predictions also indicate that ferromagnetic insulators like CrI$_3$ can similarly amplify SOC effects, along with significant magnetic exchange fields \cite{Qiao2010, PhysRevB.97.085401}. In our model, the spin-orbit coupling interaction is directly
included in the effective model of the twisted graphene bilayer, taking that the degrees
of freedom of the ferromagnet and the superconductor are integrated out. In the real device,
the spin-orbit coupling interaction will be mainly driven by the SOC of the substrate materials,
both the superconductor and ferromagnet. 
The ferromagnetic 
layer induces an out-of-plane exchange spin splitting in TBG and the superconducting 
layer is responsible for an induced s-wave superconducting order parameter.
We note that the different ingredients required to realize the Hamiltonian
above have been experimentally demonstrated independently
in a variety of van der Waals heterostructures\cite{Novoselov2004,Avsar2014,PhysRevLett.124.197401,Jo_2023,PhysRevLett.99.216802,Zhang2009,Kezilebieke2020,Bhowmik2022,Park2021}.
{ We note that an out-of-plane magnetic field will not be induced by the ferromagnetic layer,
as such field vanishes for infinite two-dimensional ferromagnets according to classical electromagnetism\cite{griffiths2017electrodynamics}}
%We note that infinite two-dimensional out-of-plane ferromagnets do not generate
%an out-of-plane magnetic field. This is a well-known result in classical electromagnetism\cite{griffiths2017electrodynamics}, and stems simply from considering that
%a plane of dipoles is equivalent to two nearby planes of opposite monopoles, and therefore their relative fields cancel each other.
%The absence of a magnetic field for an out-of-plane ferromagnetic plane is the reason why scanning SQUID microscopy or NV
%magnetometry can only stray magnetic fields at the edge of the samples, which is the only region where a finite magnetic field
%appears\cite{Song2021}.
The resulting Hamiltonian is
\begin{equation}
\H = 
\H_{0}
+
\H_J +
\H_{R}  +
\H_{\text{SC}}
\label{eq:h1}
\end{equation}
Here $H_{0}$ is the unperturbed 
tight-binding Hamiltonian for TBG, 
$H_\textbf{J}$ and $H_\textbf{SC}$ account for
the exchange fields and superconducting order 
parameter induced by proximity.
The term $H_\textbf{R}$ accounts for the Rashba SOC that results 
from the proximity-induced SOC due to mirror symmetry breaking.
We use an atomistic tight-binding Hamiltonian for TBG as 
$\H_{0}=\sum_{\langle i,j\rangle,s}t c^\dagger_{i,s}c_{j,s}+\sum_{{i,j},{s}}t^{\perp}_{i,j}c^\dagger _{i,s} c_{j,s}+ \sum_{i,s}\mu_0c^\dagger_{i,s}c_{i,s}$, 
where $s=\uparrow,\downarrow$ denotes the spin index, $i$ labels sites at position $r_{i}=(x_i,y_i,z_i)$ located at $z_i=\pm d/2$ and $\mu_0$ is an onsite potential controlling the chemical potential. The inter-layer hopping depends on the distance as $t^{\perp}_{i,j}=t_\perp[(z_i-z_j)^2/|{\bf r}_i-{\bf r}_j|^2]e^{-(|{\bf r}_i-{\bf r}_j|-d)/\lambda}$.
The typical energy scales in twisted graphene bilayers are $t=3$ eV and $t_\perp = 350$ meV\cite{RevModPhys.81.109} and we focus on a twist angle $\alpha = 1.6 ^\circ$ TBG.\footnote{Calculations are performed using a scaling
approach\cite{GonzalezArraga2017} with $t_\perp = 0.55 t$, $d=3a$,
with $a$ the C-C distance, and $\lambda_i = 8a$ }. 

The ferromagnet and the superconductor induce respectively an out-of-plane exchange field in the top layer of TBG
$\H_{J} = J \sum_{i\in T,s,s'} \sigma_z^{s,s'} c^{\dagger}_{i,s}c_{i,s'}$ 
and a pairing term in the bottom layer of TBG $\H_{\textbf{SC}} = \Delta \sum_{i \in B} c^{\dagger}_{i,\uparrow}c^{\dagger}_{i,\downarrow}+h.c.$,
Here $\sigma_z^{s,s'}$ is the spin Pauli matrix, and $i\in T$  and $i\in B$  respectively run over sites of the top and bottom layers. The mirror symmetry breaking produced by Rashba spin-orbit coupling is incorporated within both TBG 
layers and it is written as $\H_{R} =i\lambda_R  \sum_{\langle i j \rangle, s s^{'}} \mathbf{d}_{ij} \cdot \mathbf{\sigma} ^{s,s^{'}} c^{\dagger}_{i, s}c_{j,s^{'}}$, where $\mathbf{d}_{ij} = (\mathbf{r}_i - \mathbf{r}_j)\times \hat z.$ We take $J=0.018 t_\perp$, $\lambda_R=J$ and leave $\Delta$ and $\mu_0$ as tuning parameters.

The bandstructure of the unperturbed $\alpha \approx 1.6 ^\circ$  TBG  is shown in Fig.~\ref{fig:fig1}(b). Narrow bands are separated from the high energy bands by a gap at $\Gamma$, created by the moir\'e pattern. Parabolic bands are found at the $\Gamma$ and M points of the narrow bands and at the $\Gamma$ point of the high energy bands. The large moir\'e unit cell permits tuning experimentally the chemical potential to be around any of these points, doping the TBG with a gate voltage. In our setup, similar to the configuration in Refs. \cite{Cao2018}, a gate is a fundamental component to control the chemical potential of the twisted bilayer. Controlling both the interlayer bias and chemical potential of twisted multilayers is standard in these experiments, and this is the strategy we use in our model.
The addition of the exchange field, induced by the ferromagnetic layer, and the Rashba SOC  breaks the spin degeneracy, splitting the bands in Fig.~\ref{fig:fig1}(c). 
Rashba SOC  creates a momentum-dependent
splitting, while the exchange field opens up a gap
at the time-reversal invariant points, creating a pair of helical states. Except for a rigid energy shift, in the absence of superconductivity, the bandstructure of this model does not depend on doping. 
%it is important to note that 
{The exchange proximity and superconductivity
stem from the different top and bottom substrate materials. 
While both order parameters do compete, the twisted bilayer acting as a spacer prevents that the ferromagnet quenches the superconducting order. The  superconducting order $\Delta$ and exchange $J$ assumed in the model are the ones resulting after such competition is taken into account.} 
%It is important to note that the bottom ferromagnetic substrate will not quench the top superconducting substrate, given that the twisted bilayer acts as a spacer. The two proximity
%effects in the twisted bilayer are however going to compete. While such competition could be treated in a self-consistent manner, we do not expect significant changes in the effective Hamiltonian which may alter our conclusions. The resulting mean-field
%parameters will depend on fine details including the nature of the %interactions in the twisted bilayer and the coupling to the substrate.
%Candidate materials are NbSe$_2$ for SC and CrBr$_3$ for FE, and SC %has been shown to survive proximity effects in CrBr$_3$/NbSe$_2$ \cite{Kezilebieke2020}. 

\begin{figure*}
    \centering
    \includegraphics[width=\textwidth]{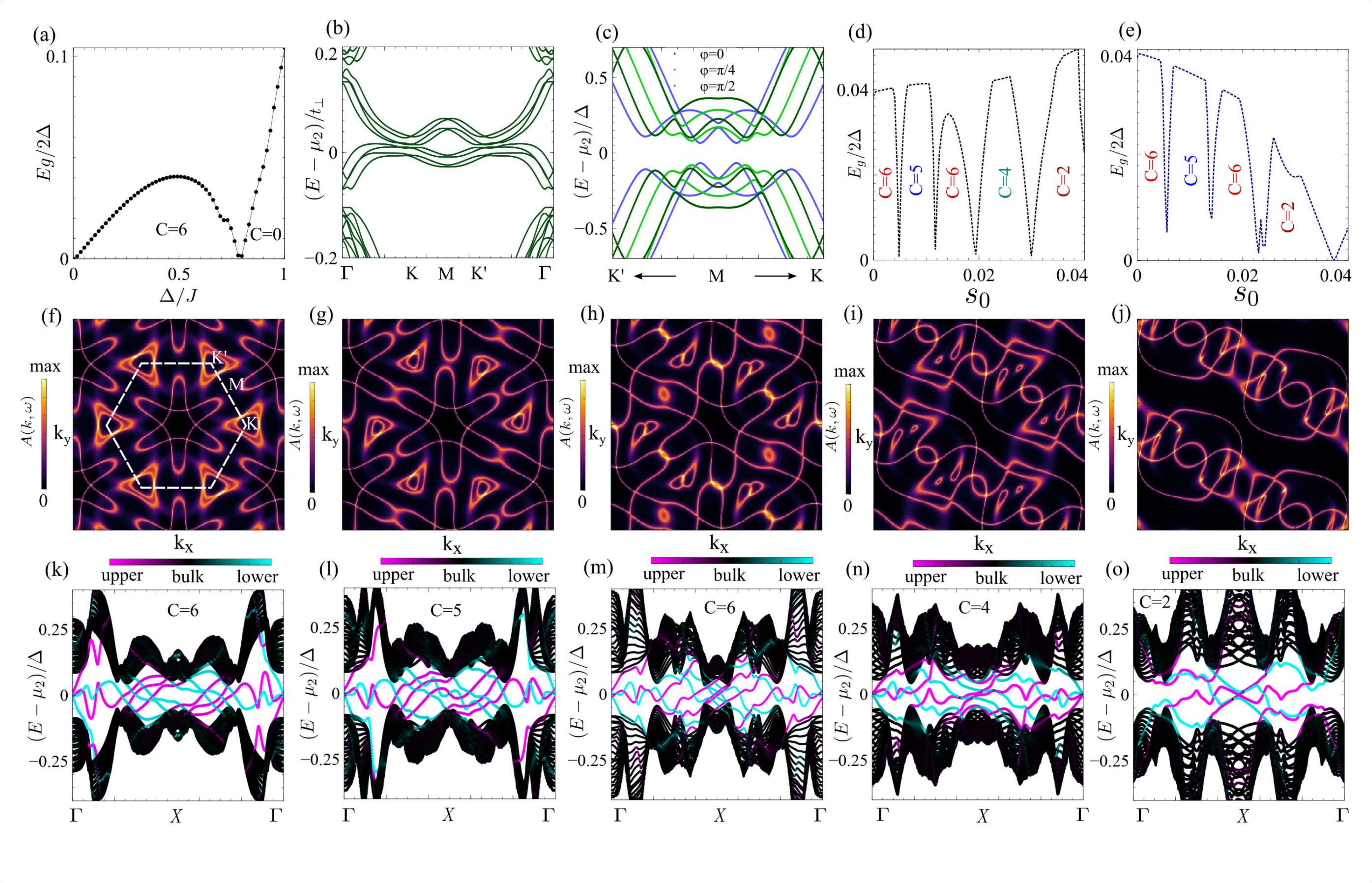}
    \caption{(a) Gap at a chemical potential $\mu_2$, close to the energy minima at the M point,  as a function of $\Delta$ in the absence of strain.     
    (b) Electronic bands of TBG under a strain of magnitude $s_0=0.025$ applied at an angle $\phi=\pi/4$ with $\Delta=0$. (c) Same as (b) close to M point with $\Delta=0.4J$ and strain  $s_0=0.015$ applied in three different directions. (d) Energy gap at $\mu_2$ as a function of the strain magnitude applied at angle $(\phi=\pi/4)$, and (e)  $\phi=\pi/2$ for $\Delta=0.4 J$. (f)-(j) Fermi surfaces at $\mu_2$ in the absence of superconductivity, $\Delta=0$, with strain applied along $\phi=\pi/4$ of magnitude $s_0=0, 0.008, 0.015, 0.025, 0.035$, respectively. (k)-(o) Energy states of finite width ribbons of width $N_r=100$ moir\'e unit cell with $\Delta=0.4 J$ as in (d) corresponding to the strain magnitudes in (f)-(j) associated to the Chern numbers in (d): (k) $C=6$, (l) $C=5$, (m) $C=6$, (n) $C=4$, (o) $C=2$ showing the emergence of chiral gapless edge states.}
    \label{fig:fig2}
\end{figure*}

\section{Induced topological superconductivity}
%\textit{Induced topological superconductivity}: 
We first investigate the existence of topological superconductivity in close proximity to the $\Gamma$ point of the lower narrow band,  at a chemical potential $\mu_1$ marked with a black dashed line in Fig.~\ref{fig:fig1}(c). 
By introducing a superconducting order parameter $\Delta$, both band inversion and an energy gap $E_g$, smaller than $\Delta$, emerge in   
the Bogoliubov-de Gennes electronic structure, 
see Fig.~\ref{fig:fig1}(d). 
We evaluate the Chern number $C$  by integrating the Berry curvature over the entire Brillouin zone and obtain $C=-2$, showing the emergence of topological superconductivity. 
This  Chern number stems from each of the two valleys, originating from the graphene K and K' points, contributing with  $C=-1$. 

A topological superconductor is expected to show $C$ chiral gapless edge states. To confirm their presence  we build a ribbon, finite along the $y$-direction and infinite in the 
$x$-direction.
The lower energy states of a wide ribbon, discrete due to the finite width, are shown in Fig.~\ref{fig:fig1} (e), and 
the color gives the projection of the states 
in the upper edge, lower edge, and bulk of the ribbon.
Bulk states, in black, are spread along the width of the ribbon. The topologically non-trivial nature of the electronic structure is confirmed by the emergence of
gapless edge states within the gap. 
two co-propagating chiral states emerge in the upper edge (magenta) and two states along the lower one (cyan),
as expected from the $C=-2$ Chern invariant. 

\section{Tuning topological superconductivity}
%{\it Tuning topological superconductivity with doping and superconducting order parameter:} 
Topological transitions can be detected by a gap closing. In Fig.~\ref{fig:fig1}(f) the gap $E_g$  is shown as a function of $\mu_0$ and $\Delta$. Regions with $C=-2$ and $C=0$ are separated by a zone with zero gap. 
Experimentally the system can be tuned through the topological transition via a gate voltage, which modifies the chemical potential. Alternatively, $\Delta$ can be changed with temperature or switching the superconducting layer.
We now move the chemical potential to $\mu_2$ close to the energy minima
at the $M$ point, blue dashed line in Fig. \ref{fig:fig1} (c). For small $\Delta$,
we observe a Chern number $C=6$ until a topological transition to a trivial state with $C=0$ takes place at larger $\Delta$ in Fig.~\ref{fig:fig2}(a). The Chern number $C=-6$ is consistent with the chiral gapless edge states found in a ribbon, see Fig.~\ref{fig:fig2}(k). The different Chern numbers at $\mu_2$ and $\mu_1$ result from their different Fermi surfaces, Fig.~\ref{fig:fig2}(f) and inset in  Fig.~\ref{fig:fig1}(c). At $\mu_2$, $C=6$ originates in the four pockets at K and K' two in each valley, and the two pockets at $\Gamma$,  one in each valley. With $\mu_3$,  in Fig. \ref{fig:fig1}c, in the high energy bands, and $\Delta=0.4 J$ we have found $C=6$ originating from a complex Fermi surface centered at $\Gamma$ and being again consistent with the ribbon edge states (not shown).

%\section{Tuning topological superconductivity with strain and displacement field}
%\textit{Tuning topological superconductivity with strain}:
Other topological superconducting states can appear
by the application of strain in the sample.
The strain is directly included in hopping terms of the Hamiltonian, such that $t_{ij}$ is renormalized as $t_{ij}(1+s_{ij})$ where $s_{ij}=s_0 (\cos{(\theta_{ij}-\phi}))^2$, $\theta_{ij}$ is the angle between sites $i$ and $j$, whereas $\phi$ is the polar angle measured with respect to the X direction parametrizing the direction of the applied strain. We assume that the same strain is applied in both graphene layers for concreteness. 
\begin{figure}[t!]
    \centering
    \includegraphics[width=\columnwidth]{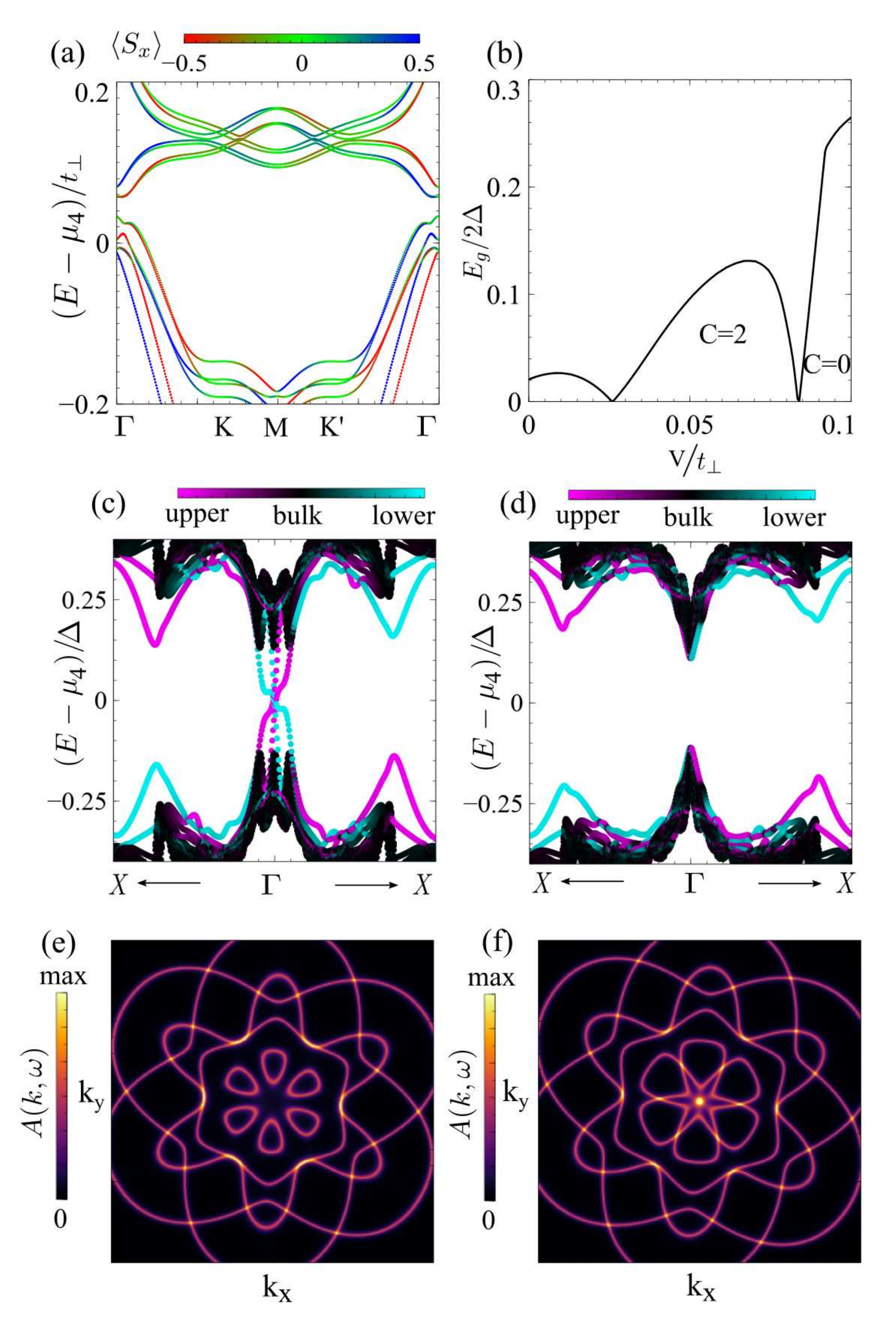}
    \caption{(a) Electronic bands centered at the higher energy bands at $\mu_4$ with doping slightly higher than at $\mu_3$, with electric bias $V=0.07t_\perp$. (b) Gap at $\mu_4$ as a function of electric bias for $\Delta=0.4J$. (c),(d) Ribbon edge states for $\mu_4$, $\Delta$ as in (b) and a bias voltage $V=0.1t_\perp$, $V=0.16t_\perp$ respectively, with $N_r=350$ moir\'e unit cells. (e), (f) Fermi surfaces at $\mu_4$ and $\Delta$ as in previous panels, respectively for the bias values  in (c) and (d).}
    \label{fig:fig3}
\end{figure} We focus on the influence of strain at $\mu_2$, close to the energy minima at the M point, originally showing $|C|=6$. The band structure subjected to a strain magnitude of $s_0=0.025$ applied along $\phi=\pi/4$, plotted in Fig.~\ref{fig:fig2}(b) in the absence of $\Delta$, shows important differences with respect to the one in Fig.~\ref{fig:fig1}(c). Topological phase transitions between different topological phases appear in Fig.~\ref{fig:fig2}(d)as a function of strain magnitude. The origin of those transitions stems from an avalanche of Lifshitz transitions driven by the strain\cite{Sinner2022}, as observed in the Fermi surfaces in Fig.~\ref{fig:fig2}(f)-(j). The phases with different Chern numbers show associated chiral modes, see  Fig.~\ref{fig:fig2}(k)-(o). 
For a fixed magnitude of the strain, the band structure depends on the direction in which it is applied, see bands in the superconducting state in Fig. \ref{fig:fig2}(c) with the strain of magnitude of $s_0=0.015$ applied under three distinct directions. Consequently, the sequence of topological transitions depends on the strain direction, which can be seen comparing Fig.~\ref{fig:fig2}(e) for $\phi=\pi/2$ with Fig. \ref{fig:fig2}(d) for $\phi=\pi/4$. 
The tunability with strain allows to engineering of a plethora of topological states in a single twisted multilayer.

%\section{Tuning topological superconductivity with displacement field}
%\textit{Tuning topological superconductivity with displacement field}:
{ Using top and bottom gates it is possible to tune selectively the doping and an out-of-plane electric field.}
We now address the tunability of the topological phases with an out-of-plane electric field. Specifically, we focus on the high energy bands obtained at doping of $\pm (4 + \epsilon)$ electron per unit cell.
An interlayer bias adds a term to the Hamilonian of the form
$\mathcal{H}_V = V \sum_{i,s} c^\dagger_{i,s} c_{i,s} \tau_{i}$, with $\tau_i= \pm 1$ for upper/lower layer.
Due to the broken inversion symmetry of structure, present already in the absence of an external electric bias, there can be an imbalance between the layers, that can be further changed with a displacement field. This allows controlling the electronic structure with an interlayer bias,  as shown in Fig.~\ref{fig:fig3}(a) in the absence of superconductivity. The normal state band structure tunability suggests that topological phase transitions driven by a bias can emerge. We now include superconducting proximity and show in Figure \ref{fig:fig3}(b) the dependence of the superconducting gap as a function of the inter-layer electric bias $V$. The gap closings driven by the bias $V$ suggest the appearance of a topological phase transition, confirmed by the calculation of the Chern numbers. Taking the specific case of $V=0.1 t_\perp, 0.16 t_\perp$, we observe that the gap closing leads to a topological transition between a topological and a trivial state with different edge channels as shown in Figures \ref{fig:fig3}(c) and \ref{fig:fig3}(d). Very much like in the case of strain and doping, the topological phase transitions are associated with Lifshitz transitions in the Fermi surface driven by the electric bias, as shown in Figures \ref{fig:fig3}(e)-(f).

{ We finally note that
the superconducting and exchange parameters in our model depend not only on the superconductivity and magnetism and top and bottom substrates but also on the coupling to the twisted bilayer. This coupling will be substrate-dependent, and therefore different encapsulations may correspond to different regimes in our topological phase diagram. In our model we have not included the super-moir\'e pattern which originates in the interplay between the moir\'e length scale of the  proximity order and the original moir\'e of the twisted bilayer. In the real system the superconducting and exchange order parameters will have a spatial modulation, dependent on the rotation angle between the twisted bilayer and the substrates. The competition of two moir\'e length scales in our system, the twisted bilayer and the exchange moir\'e, would provide a potential platform to realize topological superconductors enabled by super-moir\'e physics. While this regime is not included in the analysis of our manuscript, our results provide a starting point for such a super-moir\'e superconductor. }

\section{Conclusion}
To summarize, here we demonstrated the appearance of a family of topological superconducting states in an encapsulated twisted graphene bilayer between a ferromagnet and superconductor. We demonstrated that topological states with tunable Chern numbers appear for multilayers with chemical potential in the bottom, middle of the flat bands, and even in high energy bands. Our results show that 
high Chern number topological phases appear with the chemical potential
close to 2 electrons per moir\'e unit cell, 
leading to topological phases highly tunable via in-plane strain. Furthermore, we showed that topological states can also be tuned with an out-of-plane electric field, specifically in the higher energy bands.
For concreteness, we have considered specific configurations of strain, Rashba SOC, proximity-induced exchange, and superconducting order parameters, but
our mechanism applies to other configurations with heterostrain, or Rashba SOC at a single interface.
Importantly, stabilizing these topological states does not rely on fine-tuning the twist angle, but solely on the emergence of mini-bands in the electronic structure of the twisted multilayer. 
In summary, our results establish twisted graphene bilayers as a promising platform to realize a wide variety of topological superconducting states hosting Majorana bound states, and ultimately, establishing a potential platform for a van der Waals-based topological quantum computer. 

\textbf{Acknowledgements}:
We acknowledge the computational resources provided by
the Aalto Science-IT project,
and the financial support from the
Academy of Finland Projects Nos. 331342, 336243, and 358088,
 the Jane and Aatos Erkko Foundation,
the Jubilee Fund of the Foundation for Aalto University Science and Technology,
 and from PID2021-125343NB-100 (MCIN/AEI/FEDER,EU). 
M.K. thanks the hospitality of the Instituto de Ciencia de Materiales de Madrid. 
We thank P. Liljeroth for useful discussions.

\bibliography{biblio}

\end{document}